\documentclass[12pt]{iopart}
\usepackage{graphicx}
\usepackage[usenames]{color}
\usepackage{umoline}

\newcommand{\be}{\begin{equation}}
\newcommand{\ee}{\end{equation}}

\newcommand{\bea}{\begin{eqnarray}}
\newcommand{\eea}{\end{eqnarray}}

\newcommand{\bc}{\begin{center}}
\newcommand{\ec}{\end{center}}

\newcommand{\nat}{Nature}
\newcommand{\prl}{Phys. Rev. Lett.}
\newcommand{\pra}{Phys. Rev. A}
\newcommand{\prb}{Phys. Rev. B}

\usepackage{iopams}  
\usepackage{txfonts}  
\begin{document}

\title[Ground states and dynamics of population-imbalanced Fermi condensates in 1D]
{Ground states and dynamics of population-imbalanced Fermi condensates in one dimension}
\author{Masaki Tezuka$^1$ and Masahito Ueda$^{2,3}$}
\address{$^1$ Department of Physics, Kyoto University, Sakyo-ku, Kyoto 606-8502, Japan}
\ead{tezuka@scphys.kyoto-u.ac.jp}
\address{$^2$ Department of Physics, University of Tokyo, Bunkyo-ku, Tokyo 113-0033, Japan}
\address{$^3$ ERATO Macroscopic Quantum Control Project, JST, Tokyo 113-8656, Japan}
\ead{ueda@phys.s.u-tokyo.ac.jp}

\begin{abstract}
By using the numerically exact density-matrix renormalization group (DMRG) approach,
we investigate the ground states of harmonically trapped 
one-dimensional (1D) fermions with
population imbalance and find that the Larkin-Ovchinnikov (LO) state,
which is a condensed state of fermion pairs with nonzero center-of-mass
momentum, is realized for a wide range of parameters.
The phase diagram comprising the two phases of i) an LO state at the trap center and
a balanced condensate at the periphery and ii) an LO state at the trap center and
a pure majority component at the periphery, is obtained.
The reduced two-body density matrix indicates that most of the minority
atoms contribute to the LO-type quasi-condensate.
With the time-dependent DMRG, we also investigate the real-time dynamics of 
a system of 1D fermions in response to a spin-flip excitation.
\end{abstract}

\vspace{2pc}
%\noindent{\it Keywords}: Cold atoms, superfluidity, FFLO, real-time dynamics, density-matrix renormalization group
\submitto{\NJP}
\maketitle

\setcounter{tocdepth}{2}
\tableofcontents
\markboth{}{}

\section{Introduction}

One-dimensional atomic gases have been vigorously investigated
over the last decade.
By using a strongly anisotopic Ioffe-Pritchard-type magnetic trap
\cite{PhysRevLett.87.130402} or pairs of counterpropagating laser beams
\cite{2001PhRvL..87p0405G}, elongated traps of atoms have been realized.
Tens to hundreds of ultracold atoms have been trapped
in such an elongated potential
that the maximum kinetic energy of the atoms are much smaller than the energy
separation between the two lowest transverse levels of the potential
\cite{2004Natur.429..277P, PhysRevLett.94.210401, 2006Natur.440..900K, 2008NatPh...4..489H, 2009arXiv0912.0092L}.
Thus the physics of the system becomes essentially
one-dimensional \cite{PhysRevA.54.656, PhysRevLett.81.938}.
For the case of bosons, a strongly interacting Tonks-Girardeau regime was
realized \cite{2004Natur.429..277P};
no equilibration of an integrable system --- ``quantum 
Newton's cradle'' --- was demonstrated \cite{2006Natur.440..900K};
and the crossover from thermal to quantum noise was also observed \cite{2008NatPh...4..489H}.

For the case of fermions,
the 1D atomic gas allows one to study an analogue of a system of electrons in solids.
The numbers of atoms in trappable hyperfine states can be controlled individually,
so that by populating two hyperfine states simultaneously, one can study effects
of population imbalance on the properties of one-dimensional systems.
While a true long-range order is forbidden in 1D even at $T=0$ by the
Hohenberg-Mermin-Wagner theorem,
(quasi) condensation and superfluidity are possible in a finite-sized,
trapped system. The search for 
Fulde-Ferrell-Larkin-Ovchinnikov (FFLO) \cite{PhysRev.135.A550, larkin:1964zz}
pairing states has been of major experimental interest in 1D fermionic systems.

The confinement-induced two-particle bound state was observed 
\cite{PhysRevLett.94.210401} in an array of quasi one-dimensional
tubes with equal populations of two hyperfine states,
validating the 1D treatment of these tubes.
More recently, the Rice group \cite{2009arXiv0912.0092L}
has realized quasi one-dimensional systems of population-imbalanced cold atoms 
and measured the density profile for each of the hyperfine species.

In this article we first report our study on the ground state
of a system of population-imbalanced fermions in harmonic traps \cite{tezuka:110403}.
The system has also been numerically studied by
using the density-matrix renormalization group
(DMRG) \cite{feiguin:220508, 2008PhRvB..77x5105R,
2008PhRvA..77e3614M, 2008PhRvA..78a3637L, 2008PhRvB..78w5117M,
PhysRevA.81.023629, 2010arXiv1001.4720H},
Bogoliubov-de Gennes (BdG) equations \cite{2008PhRvL.101l0404B, 2008PhRvA..78b3601L},
and Quantum Monte Carlo (QMC) method \cite{batrouni:116405, 2008PhRvA..78c3607C, 2009NJPh...11e5041R}.
The phase diagram including an FFLO superconductor region for a 1D homogeneous system
was obtained by bosonization \cite{PhysRevB.63.140511}.
Studies on trapped systems based on the exact solution of the Gaudin-Yang model
were carried out in Refs. \cite{orso:070402, hu:070403, 2008PhRvA..77c3604G, 2009PhRvA..79d1603K, 2010JLTP..158...36Z}.
Other related studies include 
the cases with unequal (effective) masses and the balanced or
imbalanced numbers of atoms
\cite{2009PhRvA..79e1604W, 2009PhRvA..80a1601G, 0295-5075-86-4-47006, 2009arXiv0912.4205B};
an angular FFLO state in a toroidal trap \cite{2009PhRvB..80v0510Y};
effects of particle-correlated hopping in an optical lattice \cite{2009PhRvA..79d3612W};
imbalanced Fermi gases on two-leg ladders \cite{2009PhRvL.102g6403F} and
two-dimensional optical lattices \cite{2008PhRvA..78a1603I};
composite particles of more than two particles \cite{2009PhRvL.103u5301B};
universality and Efimov states in systems where two or more species of atoms are 
confined in traps having
different dimensions \cite{nishida:170401, nishida:060701};
transport through a Y-junction and ring-geometry systems \cite{2008PhRvL.100n0402T}.

We note that, for the ground state that is symmetric under time reversal,
the condensate cannot be of a Fulde-Ferrell type, because
the Fulde-Ferrell state is described by a pair correlation function that is
a complex-valued function of the spatial distance.
On the other hand, the Larkin-Ovchinnikov state is not forbidden by the time-reversal
symmetry because the correlation functions can be real-valued.

Here we numerically study the ground-state wavefunction of the trapped system by
discretising the system and the DMRG
\cite{PhysRevLett.69.2863, PhysRevB.48.10345, RevModPhys.77.259} method.
We calculate the density distribution of individual spin components in the trap,
which is a physical quantity measured in \textit{in-situ} light absorption
or phase shift measurements
\cite{M.R.Andrews07051996, 1998Natur.392..151I, PhysRevLett.96.130403},
and show that the density-difference distribution exhibits oscillations which are
the hallmark of the Larkin-Ovchinnikov state \cite{PhysRevB.30.122}.

In the second part of the paper,
we study the real-time evolution of a system of trapped Fermi atoms
after a sudden spin flip.
Because of the large interparticle distance and low speed of the atoms,
the dynamics is extremely slow as compared to electrons in solid state physics.
Moreover, the atoms are trapped in an optical and/or magneto-optical potential
in the vacuum,
so that the dynamics can be optically observed in real time.
As the state of the atoms can be manipulated by microwaves,
it is of interest to study
the response of the 1D ground state to a change in population.

\section{The ground state}

First, we study the ground state of the trapped, population imbalanced
fermionic system in 1D within the single-channel model.
The effect of a molecular state in the BCS-BEC crossover in 1D systems
\cite{PhysRevA.71.033630}
has also been studied in population imbalanced systems
\cite{PhysRevA.81.023629, PhysRevA.81.033628},
but here we focus on the case in which the contribution of the molecular state can be neglected.

The interaction between ultracold atoms is described by the $s$-wave scattering
which can be approximated by the Lee-Huang-Yang pseudopotential
\cite{Huang}
\be
U(r) = g_s^\mathrm{3D}\delta(\vec{r})\frac{\partial}{\partial \vec{r}}(r\cdot\,),
\ee
with the bare 3D coupling constant $g_s^\mathrm{3D}$.
The relation between $g_s^\mathrm{3D}$ and the $s$-wave scattering length
$a_s^\mathrm{3D}$ is given by $g_s^\mathrm{3D} = 2\pi\hbar a_s^\mathrm{3D}/\mu$,
where $\mu$ is the reduced mass.

The relation between $a_s^\mathrm{3D}$
and the scattering length in a 1D trap $a_s^\mathrm{1D}$
has been given in \cite{PhysRevLett.81.938},
for the case in which the transverse confinement can be approximated
by a 2D harmonic potential with (angular) frequency $\omega_\perp$.
Provided that the size of the ground state of the transverse confinement
$a_\perp \equiv (\hbar/\mu \omega_\perp)^{1/2}$
is much larger than an effective range $r_0$ of the bare potential,
$a_s^\mathrm{1D}$ is given by
\be
a_s^\mathrm{1D} = -\frac{a_\perp^2}{2a_s^\mathrm{3D}}\left(1-C\frac{a_s^\mathrm{3D}}{a_\perp}\right),
\ee
with $C=1.4603\ldots$.
The 1D scattering amplitude for the relative wavenumber $k_z$ is expressed as
\be
f_\mathrm{even}(k_z)= -\frac{1}{1+ik_za_s^\mathrm{1D}+\Or\left((k_za_\perp)^2\right)},
\ee
which, in the low-energy limit ($|k_za_\perp|\ll1$),
reduces to the scattering amplitude derived from the one-dimensional $\delta$ potential
\be
U^\mathrm{1D}(z) = g_s^\mathrm{1D}\delta(z)
\ee
with $g_s^\mathrm{1D} = -\hbar^2/(\mu a_s^\mathrm{1D})$ \cite{PhysRevLett.81.938}.

When we have two fermionic species with the equal mass $m_0$,
the reduced mass is $\mu\equiv m_0/2$.
The two hyperfine states can be regarded as a pseudo spin-$1/2$ system with spin
up ($|\uparrow\rangle$) and down ($|\downarrow\rangle$) states.
We assume that the up-spin state is the majority state.
We start with an effective 1D Hamiltonian for the continuum,
\be
\hat \mathcal{H}_\mathrm{cont} \equiv \sum_{a,\sigma}\left(
-\frac{\hbar^2}{2m}
\frac{\partial^2}{\partial z_{a,\sigma}^2} + V(z_{a,\sigma})\right)
+ \sum_{a,a'}g^\mathrm{1D}\delta(z_{a,\uparrow}-z_{a',\downarrow}),
\ee
where $m$ is the atomic mass, $V(z) \equiv \kappa z^2/2$ is the harmonic potential
in the $z$ direction
and $g$ is the coupling constant.
We discretize this Hamiltonian
to perform calculations in the DMRG framework.

We take a characteristic size of the potential, $l$,
as the unit of length, and denote
a characteristic trap depth by $A$: $V(\pm l)=A$.
We discretize the system by introducing a lattice of $L$ sites with lattice constant
$d\equiv 2l/L$.
The transfer amplitude between the neighboring sites is chosen to be
$J\equiv \hbar^2/(2md^2)$.
The band dispersion, $E(k) = 2J(1-\cos kd) = J(kd)^2 + O(k^4)$,
where $k=p/\hbar$ is the wave number of the atom with momentum $p$,
reproduces the energy dispersion of the atom in the free space, $E(k) = p^2/(2m) = \hbar^2k^2/(2m)$,
in the vanishing limit of the filling factor ($L\rightarrow\infty$).
The interaction $g^\mathrm{1D}\delta(z_{a,\uparrow}-z_{a',\downarrow})$ is approximated
by an on-site interaction,
$U\sum_{i=0}^{L-1} \hat n_{i,\uparrow} \hat n_{i,\downarrow}$,
where $i$ runs over all lattice sites $0,1,\ldots,(L-1)$,
with coupling constant $U\equiv g^\mathrm{1D}/d$.
In the following we take $\hbar=m=1$.
We consider the case of negative $U$,
which corresponds to a positive $a_s^\mathrm{1D}$
and attractive interaction between atoms with opposite spins.

\subsection{Relations between the experimental and discretized model parameters}

Here, we describe how the various parameters introduced above can be related to the experimental parameters.
For the ${}^{40}\mathrm{K}$ atoms with $\omega_r = 2\pi \times 69~\mathrm{kHz}$
and $\omega_z = 2\pi \times 256~\mathrm{Hz}$
\cite{PhysRevLett.94.210401},
$a_\perp = 86~\mathrm{nm}$, and
at a 3D Feshbach resonance ($a_s^\mathrm{3D}\rightarrow\infty$),
we have $a_s^\mathrm{1D} = 62.8~\mathrm{nm}$.
A typical size of the axial ground-state wave function of a single atom in this trap is
$a_z \equiv (\hbar / \mu \omega_z)^{1/2} = 1.41~\mu\mathrm{m}.$
As the (half-)width of the simulated region of the harmonic trap, we take
$l=6.28~\mu\mathrm{m}$.
The length will be measured in units of $l$ below.
Now, $J=\hbar^2/\left(2m(2l/L)^2\right) = \hbar^2L^2 / (8m_0 l^2)$,
so that $A/J = (8A m_0 l^2 / \hbar^2)L^{-2}$ and
$U/J = -\hbar^2/(m_0 a_s^\mathrm{1D} d/2) / J = -(8l/a_s^\mathrm{1D})/L$.
We choose 
$A = k_\mathrm{B}\times 246~\mathrm{nK}$,
so that $A/J = 6400 L^{-2}$ and $U/J = -800L^{-1}$, unless otherwise stated.

If there are $N$ atoms per spin, the Fermi energy for the noninteracting case is
$E_\mathrm{F}=N\hbar\omega_z$. For the same number of atoms per spin, the dimensionless 
strength of interaction $\xi$ at the Fermi level is
\be
\xi^0(N) = \left(k_\mathrm{F}a_s^\mathrm{1D}\right)^{-1}
= \frac{\sqrt{\hbar/(2m_0\omega_z)}}{a_s^\mathrm{1D}}N^{-1/2}
= 11.2 N^{-1/2}.
\ee
Similarly, for $n$ atoms per spin per unit length $l$, the Fermi wave vector is
$k_\mathrm{F} = \pi n/l$, and we obtain
\be
\xi(n) = \left(k_\mathrm{F}a_s^\mathrm{1D}\right)^{-1}
= (l/\pi a_s^\mathrm{1D})n^{-1}
= 31.8 n^{-1}.
\ee

\subsection{Method: Density-matrix renormalization group}

After the discretization described above,
the Hamiltonian is given by
\be
\hat \mathcal{H} = -J\sum_\sigma\sum_{i=1}^{L-1}
(\hat c^\dag_{i,\sigma} \hat c_{i-1,\sigma} + \mathrm{h.c.})
+ U\sum_{i=0}^{L-1} \hat n_{i,\uparrow} \hat n_{i,\downarrow}
+ \sum_{i=0}^{L-1} V(i) (\hat n_{i,\uparrow} + \hat n_{i,\downarrow}),
\label{eqn:Hamiltonian}
\ee
where $V(i) = \kappa (i-C)^2$ with $\kappa \equiv 4A/L^2$ and $C \equiv (L-1)/2$.
Note that $A/J\propto L^{-2}$ (thus $\kappa/J\propto L^{-4}$)
and $U/J\propto L^{-1}$.
We use DMRG to calculate the ground state of 
the Hamiltonian (\ref{eqn:Hamiltonian}) within the 
particle-number sector of $N_\uparrow$ and $N_\downarrow$ atoms
in spin-up and spin-down states, respectively.
The population imbalance parameter is defined as
$P\equiv (N_\uparrow-N_\downarrow)/N_\mathrm{total}$,
where $N_\mathrm{total}\equiv N_\uparrow + N_\downarrow$.
The Fermi momentum at the trap center is calculated
from the averaged density
$n_\sigma \equiv \overline{ \langle \hat n_{i,\sigma} / d \rangle }$ as
$k_{\mathrm{F}\sigma}\equiv \pi n_\sigma$,
where the overbar denotes the average over $0.1L - 0.2L$ neighboring sites.

Typically $m=300$ states per block are retained and 4-5 finite system
iterations are required for DMRG convergence after the standard infinite system
initialization or the recursive sweep method initialization \cite{JPSJ.76.053001}.
We target the ground state with $N_\uparrow$ and $N_\downarrow$ atoms
in spin-up and spin-down states
as well as several states with a few atoms added or removed,
in order to improve the convergence of the pair correlation.
While we do not assume the reflection symmetry of the wavefunction in the
finite system iteration process, the reflection symmetry emerges spontaneously
for the ground state wavefunction for any of the set of parameters studied.
The sum of the eigenvalues of the reduced density matrix
for the discarded eigenstates in each DMRG step is below $2\times 10^{-5}$
for $N=64$ and $L=320$; it is below $5\times 10^{-7}$ for $N=40$ and $L=200$
if we target the ground state alone.

\subsection{Results: the Larkin-Ovchinnikov quasi-condensate}
\subsubsection{Pair correlation function}
\begin{figure}
%\bc
\hspace{\fill}
\includegraphics[width=12cm]{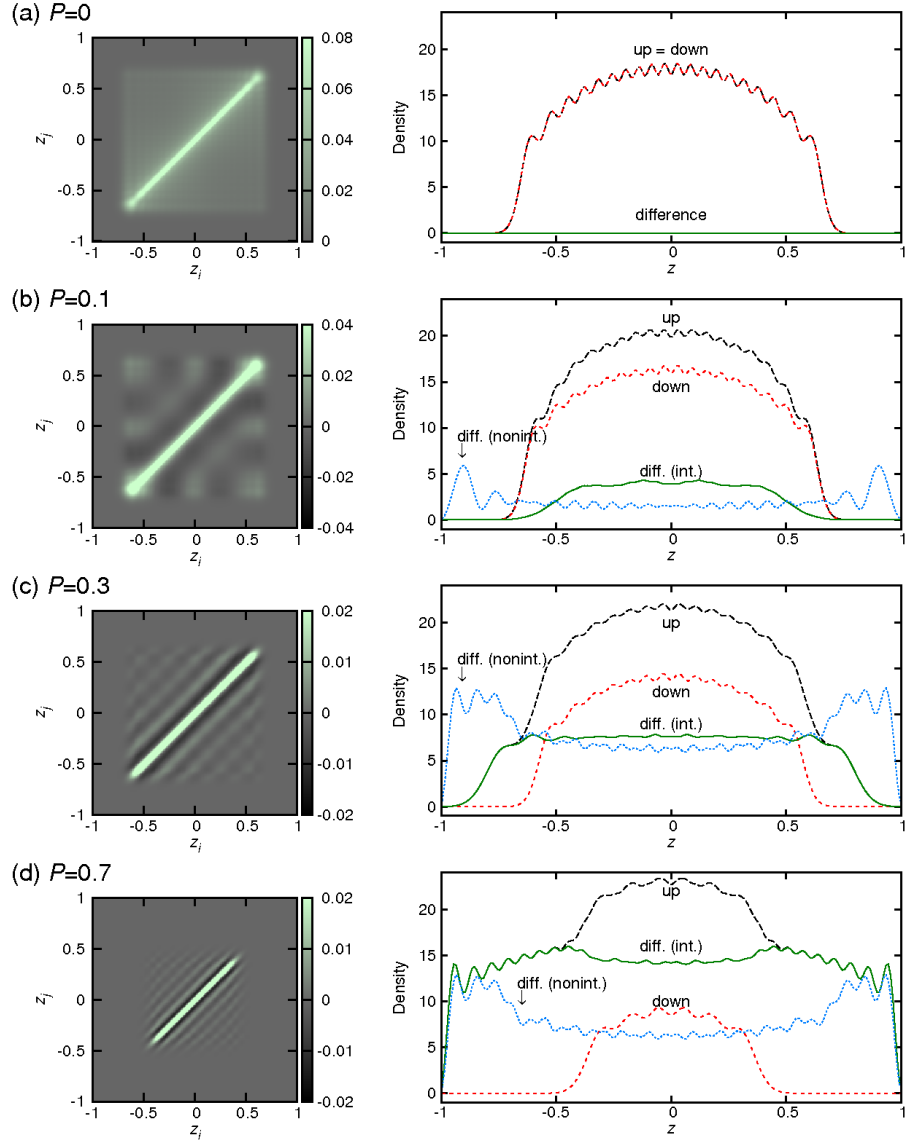}%{CorrDensity.eps}
%\ec
\caption{
Left column: 
pair correlation function (\ref{eqn:pairCorrelation})
for $L=200$, $N=40$, $A/J=0.16=6400/L^2$, $U/J=-4=-800/L$, and $P=0, 0.1, 0.3, 0.7$.
Right column: the corresponding density distribution functions for majority (up),
minority (down), and their difference (diff.) with (int.) and without (nonint.) interaction.
(a)(b): reproduced from \cite{tezuka:110403} with modified color codes.}
\label{fig:Cor00_10_30_70}
\end{figure}

We first examine the on-site pair correlation function defined as
\be
O_\mathrm{on-site}(z_i, z_j)
\equiv \langle \Psi_0|\hat c^\dag_{i, \uparrow} \hat c^\dag_{i, \downarrow}
\hat c_{j, \uparrow} \hat c_{j, \downarrow}|\Psi_0\rangle,
\label{eqn:pairCorrelation}
\ee
where
$O_\mathrm{on-site}(z_j, z_j) = \langle n_{j,\uparrow} n_{j,\downarrow}\rangle$
gives the double occupancy ratio. 
The left column of figure \ref{fig:Cor00_10_30_70} shows
$O_\mathrm{on-site}(z_j, z_j)$
in color-coded two-dimensional plots for several different values of the imbalance
parameter $P$.
For $P=0$, the pair correlation is maximal for $z_i = z_j$
with the double occupancy per site,
decreases with increasing $|z_i-z_j|$, and precipitously
drops to zero where the atomic density, shown in the right column, vanishes.

As we increase $P$, the correlation function starts to oscillate.
The zeros of the pair correlation $O_\mathrm{on-site}(z, z_j)$ for fixed $z_j$ appear
almost periodically, except for the region $|z-z_j|\ll l$, where $O_\mathrm{on-site}$ is
positive and large.
The sign change of the pair correlation reflects that of the pair amplitude.
We note that the whole region
with a significant density of minority atoms is strongly correlated inside the 1D trap.

\subsubsection{Density distribution: Oscillation of density difference\\}

\begin{figure}
%\bc
\hspace{\fill}
\includegraphics[width=12cm]{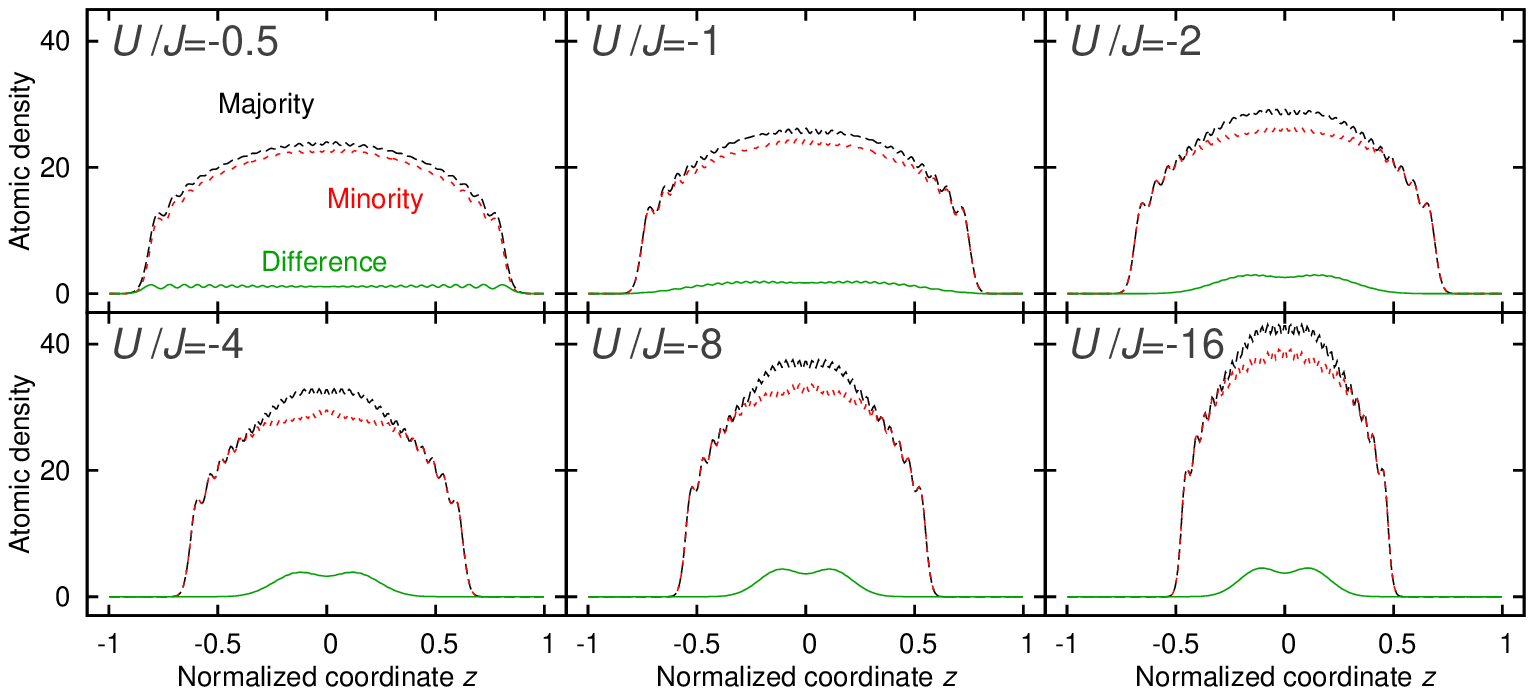}%{MajMinDif.eps}
%\ec
\caption{%(Color online)
Atomic densities of the majority and minority atoms and their difference
for $L=320$, $N=64$, $A/J=0.2 = 20480/L^2$, $P=0.03$, and $U/J=-0.5, -1, -2, -4, -8, -16$.}
\label{fig:MajMinDif}
\end{figure}

\begin{figure}
%\bc
\hspace{\fill}
\includegraphics[width=11cm]{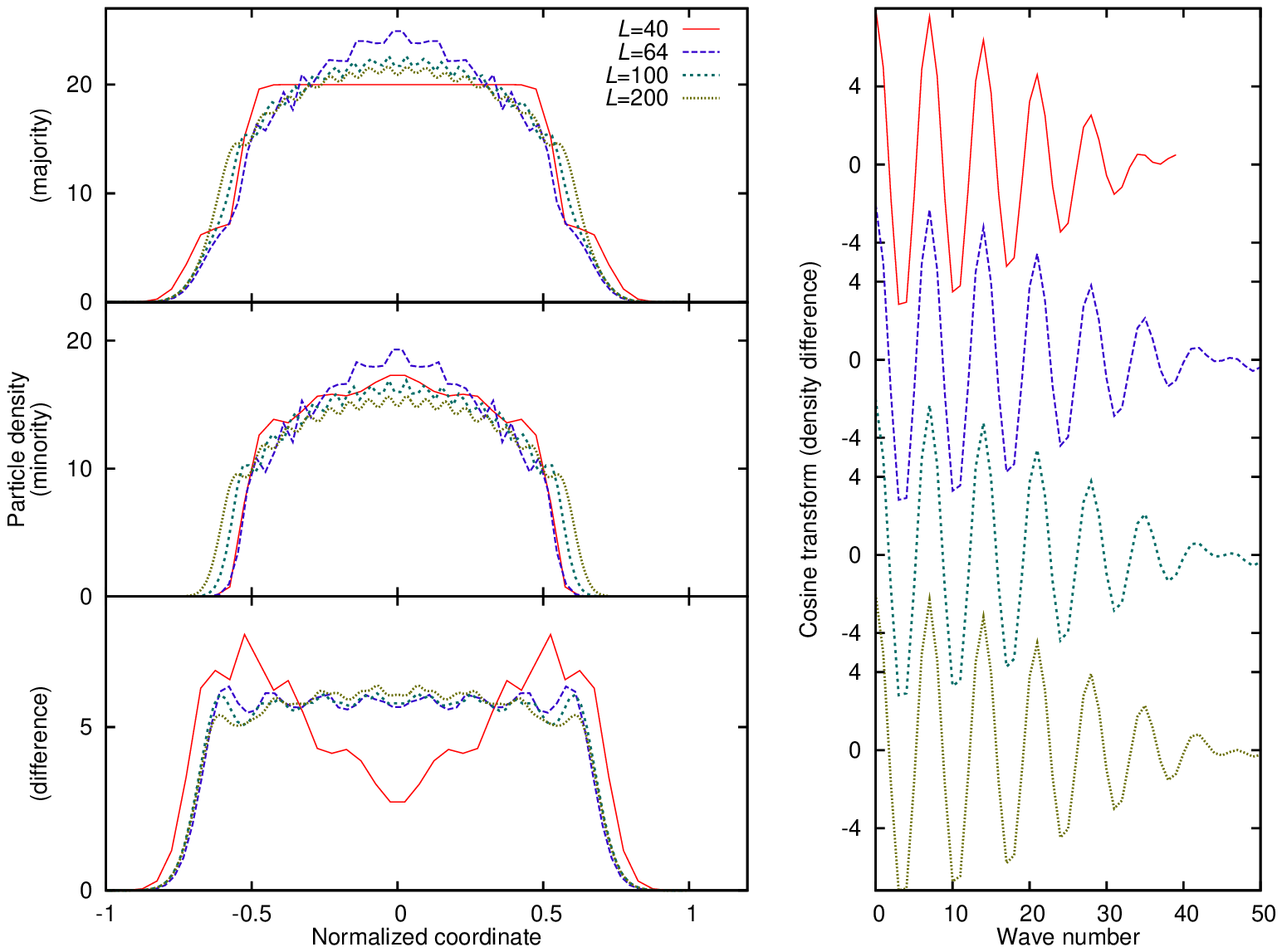}%{2416.eps}
%\ec
\caption{%(Color online)
Left column: Density distributions of majority and minority atoms and their difference plotted
against the normalized coordinate in the harmonically trapped system of fermions
with $(N_\uparrow, N_\downarrow) = (24, 16)$ atoms and for
different values of $L$. $A/J = 6400/L^2$ and $U/J=-800/L$.
Right column: Fourier (cosine) transformations of the density distribution
plotted against the wave number.}
\label{fig:L}
\end{figure}

Next, we examine the density distributions of the spin-up and spin-down atoms.
For the non-interacting system, the extensions of the majority and minority
distributions are determined by those of the $N_\uparrow$-th and
$N_\downarrow$-th eigenstates of the trap, respectively.
As $|U|$ is increased, the attractive interaction causes the atomic distributions
to shrink towards the trap center, as shown in figure \ref{fig:MajMinDif}.
We also find that the density difference
$\langle n_{i,\uparrow}-n_{i,\downarrow} \rangle/d$
is initially spread all over the region with a nonzero atomic distribution,
but that it shrinks more rapidly than the atomic distribution for
sufficiently large $|U|$.
The oscillation is independent of the lattice constant $d$ for large $L$,
as can be seen in figure \ref{fig:L}.
The peaks in the frequency spectrum $D(q)$ of the density difference,
calculated as
\be
D(q)\equiv \sum_i \cos(q z_i) (n_\uparrow - n_\downarrow),
\ee
and shown in the right column of figure \ref{fig:L}, do not
show a significant shift upon increasing the
system size.

\begin{figure}
%\bc
\hspace{\fill}
\includegraphics[width=11cm]{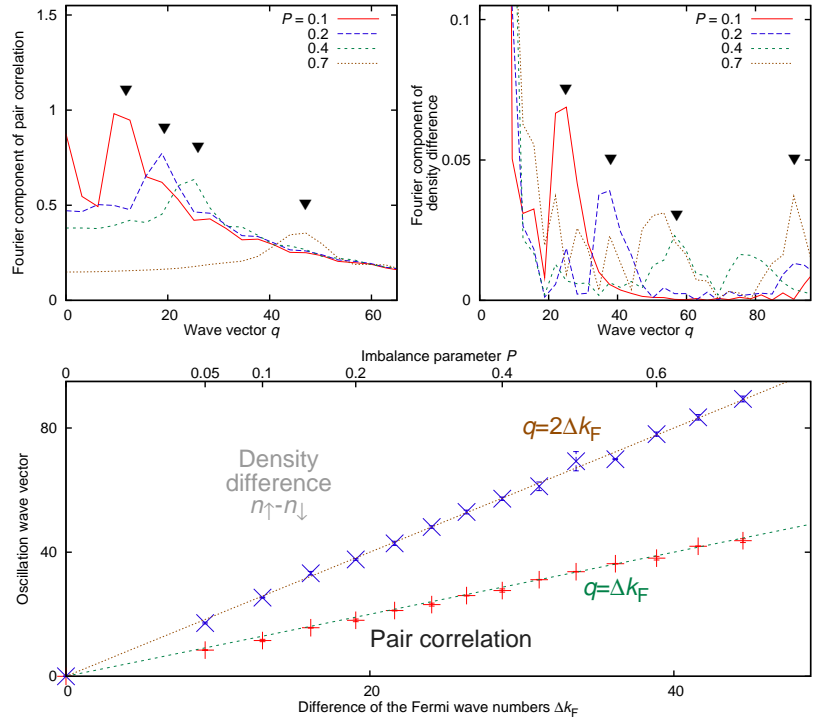}
%\ec
\caption{%(Color online)
Upper : 
Fourier components of the pair correlation function (left)
and the density difference (right; see the main text) plotted against the wave vector.
Lower (reproduced from \cite{tezuka:110403} with modified color codes):
The wave numbers of the oscillations of the pair correlation function and
the density difference plotted against the difference of the Fermi wave numbers
at the trap center.
The values of the parameters are $L=200$, $N=40$, $A/J=0.16=6400/L^2$, $U/J=-4=800/L$.
}
\label{fig:kPairDiff}
\end{figure}

From the spin-dependent Fermi momentum at the trap center,
$k_{\mathrm{F}\sigma}$, we obtain the difference of the Fermi momenta,
$q\equiv k_{\mathrm{F}\uparrow} - k_{\mathrm{F}\downarrow}$.
In the LO state, an up-spin atom close to the up-spin Fermi point (in 1D) $\pm k_{\mathrm{F}\uparrow}$
and a down-spin atom close to the down-spin Fermi point $\mp k_{\mathrm{F}\downarrow}$
form a pair; therefore the total momentum of the pair must be close to $\pm q$.
In figure \ref{fig:kPairDiff} we plot the cosine transformations of the
pair correlation $O_\mathrm{on-site}(z,z_{L/2})$, where
$z_{L/2} = 1/L$ is the location of one of the two sites closest to the trap center $z=0$,
and the density difference $2S_z(z) \equiv n_\uparrow(z) - n_\downarrow(z)$
against the wave vector.
To clearly identify oscillations close to the trap center,
the density difference has been multiplied
by a slowly varying numerical factor, $\zeta(z)\equiv \exp{(-6z^2)}$, before Fourier transformation.
The Fourier components of the pair correlation and density difference at wave vector $q$ are defined as
\be
\tilde O_\mathrm{on-site}(q)\equiv \left|\sum_i \cos(kz_i) O_\mathrm{on-site}(z_i,z_{L/2})\right|,
\ee
and
\be
\tilde n_\mathrm{diff}(q)\equiv \left|\sum_i \cos(kz_i) \zeta(z) 2S_z(z_i)\right|.
\ee
As shown in figure \ref{fig:kPairDiff}, the oscillation wave vector of the
pair correlation function is always close to $q$.
This is consistent with studies of population imbalance in the optical lattice systems,
namely, the pair momentum distribution function calculated in \cite{feiguin:220508},
Fourier transform of the pairing correlation \cite{2008PhRvB..77x5105R},
on-site pair amplitude \cite{2008PhRvA..77e3614M},
spatial noise correlations \cite{2008PhRvA..78a3637L}, and
radio-frequency spectroscopy calculation \cite{2008PhRvL.101l0404B},
as well as the pair momentum distribution calculated
in \cite{batrouni:116405, 2008PhRvA..78c3607C} for the continuum.
The oscillation wave vector of the density difference is close to $2q$,
as expected in the LO state \cite{PhysRevB.30.122}.
This result is consistent with
the spin-spin correlation function studied in \cite{2009NJPh...11e5041R}
and distribution of spin polarization studied by solving
BdG equations in \cite{2008PhRvA..78b3601L}.

\subsubsection{Phase diagram: Parameter region with phase separation}

We find the following two phases:
i) the spin-down (minority) atoms occupy a smaller region compared with the spin-up (majority) atoms
and the periphery of the system is occupied by the spin-up normal component alone,
and 
ii) the densities of spin-up and spin-down atoms are different at the trap center but
become almost equal over a finite width close to the periphery.
These two situations can be thought of as two ``phases'' of the system.
The transition between the two phases occurs when
the trap is filled with the LO condensate.
Such a phase transition was predicted in studies in which the exact solution of the Gaudin-Yang
model without a trapping potential was plugged into the trapped system
via the local-density approximation (LDA) \cite{orso:070402, hu:070403}.

For a given scattering length $a$ or coupling constant $g$,
case i) is realized for greater spin polarization $P$, where the condensate can no longer
hold all the majority atoms and the system gains energy by pushing some of the majority
component to the edge of the trap.
This is because the condensation energy of the remaining LO condensate increases
by a decrease in the Fermi wave-number difference and the number of
nodes of the pairing potential.

\begin{figure}
%\bc
\hspace{\fill}
\includegraphics[width=11cm]{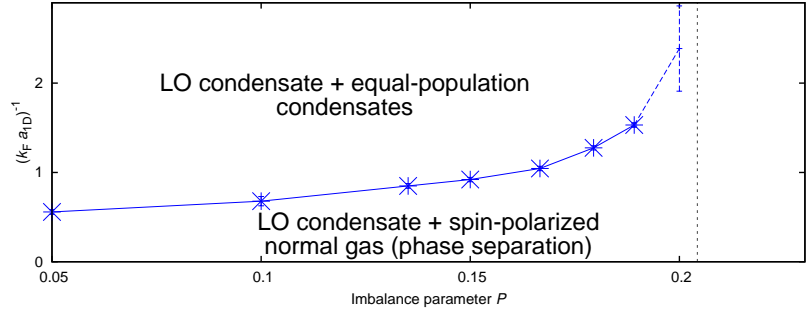}%{PhaseDiagram.eps}
%\ec
\caption{Phase diagram of the trapped population-imbalanced
Fermi gases: i) an LO condensate at the trap center and spin-polarized normal gases
that include only the majority component
at the edges of the trap, and ii) an LO condensate at the trap center and equal-population condensates at the edges of the trap.}
\label{fig:phase_diagram}
\end{figure}

Because the coupling constant $|g|$ increases with increasing $|U|$, which corresponds
to shorter $|a|$ in 1D, the critical point of phase separation shifts toward greater values of $P$,
as shown in figure \ref{fig:phase_diagram}.
The phase separation curve, however, does not exceed $P\simeq 0.21$.

\subsubsection{Effect of asymmetric potential: Accuracy of local density approximation\\}
\begin{figure}
%\bc
\hspace{\fill}
\includegraphics[width=12cm]{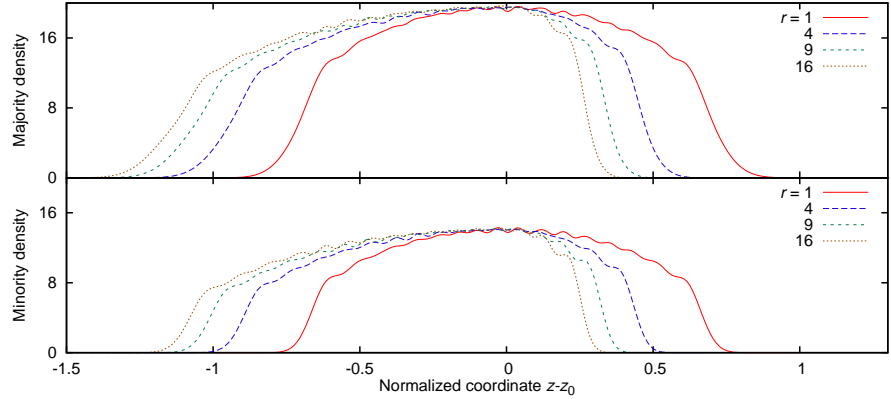}%{1DFFLO/cv2/asymtrap/NvsZ.eps}
%\ec
\caption{%(Color online)
Density distributions for spin-up (top) and spin-down (bottom) atoms
plotted
against the normalized trap coordinate shifted by $z_0$,
where $z_0$ is the solution of $V(z_0)=0$,
so that the $z-z_0=0$ corresponds to the trap minimum.
The parameters used are $L=200$, $N=40$, $P=0.2$, $A/J=6400/L^2=0.16$, and $U/J=-500/L=-2.5$.
}
\label{fig:NvsZ}
\end{figure}

\begin{figure}
%\bc
\hspace{\fill}
\includegraphics[width=12cm]{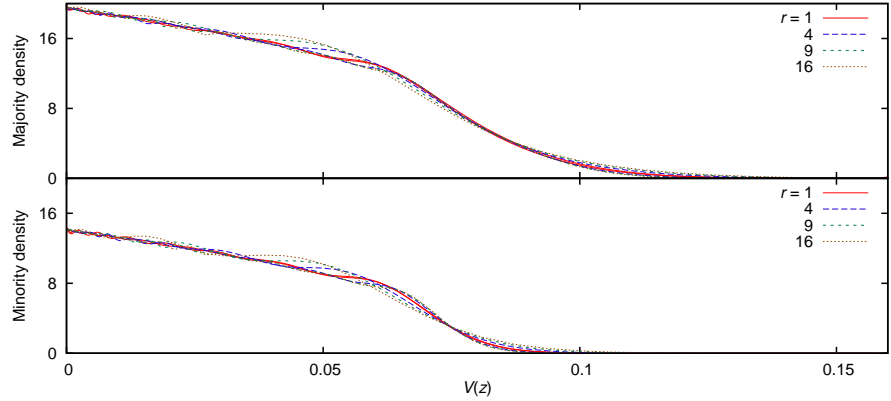}%{1DFFLO/cv2/asymtrap/VvsZ.eps}
%\ec
\caption{%(Color online)
Density distributions for spin-up (top) and spin-down (bottom) atoms for
$L=200$, $N=40$, $P=0.2$, $A/J=6400/L^2=0.16$, and $U/J-500/L=-2.5$,
plotted against the site energy level $V(z)$.
Here $r\equiv \kappa_+ / \kappa_-$ gives the ratio between the potential slopes
shown in (\ref{eqn:asymPot}).
}
\label{fig:VvsZ}
\end{figure}

To study the robustness of the Larkin-Ovchinnikov state with respect to the trap geometry
and the slope of the trap,
we examine the ground state of the system for a spatially asymmetric potential,
\be
V_\mathrm{asym}(z_i) \equiv \left\{
\begin{array}{cc}
\kappa_+(i-C)^2 & (i>C) \\
\kappa_-(i-C)^2 & (i<C)
\end{array}
\right.,
\label{eqn:asymPot}
\ee
where $0<C<L-1$, $\kappa_+ = A/C^2$ and $\kappa_- = A/(L-1-C)^2$.
The bottom of the trap is located at $i=C\equiv L\sqrt{r}/(\sqrt{r}+1)$.
Let $r \equiv \kappa_+/\kappa_-$ be the ratio of the spring constants of the
harmonic trap; this ratio equals the ratio between the potential slopes in the right and
left sides of the bottom of the trap.
Figure \ref{fig:NvsZ} shows the density distribution against the shifted trap coordinate
$z-z_0$, where $z_0 = 2C/L-1=(\sqrt{r}-1)/(\sqrt{r}+1)$ and $V(z_0)=0$.
Figure \ref{fig:VvsZ} shows the same density distribution
plotted against the site energy level $V(z)$.

\begin{figure}
%\bc
\hspace{\fill}
\includegraphics[width=12cm]{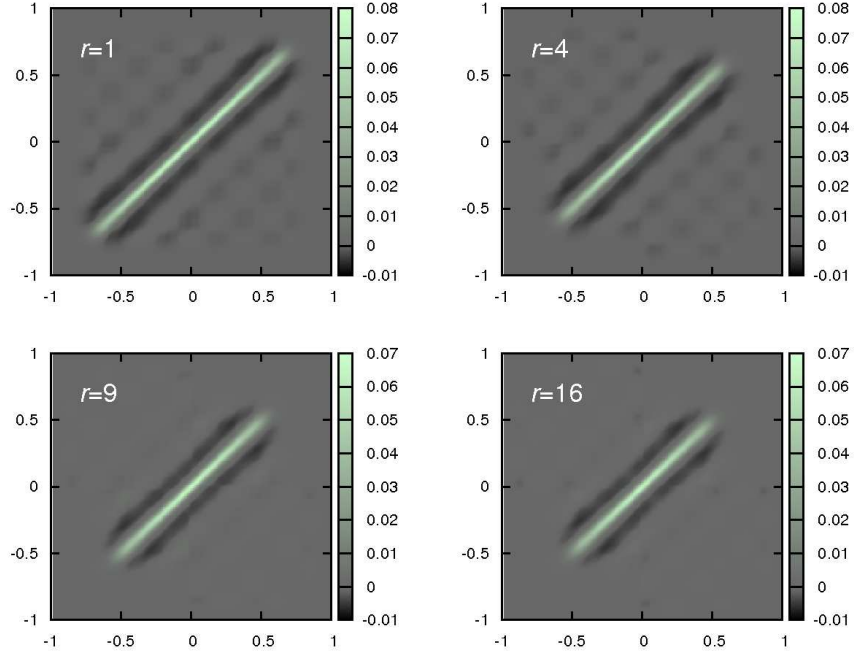}%{U24D16.eps}
%\ec
\caption{%(Color online)
On-site pair correlation for $r=1, 4, 9, 16$,
$L=200$, $N=40$, $P=0.2$, $A/J=6400/L^2=0.16$, and $U/J=-500/L=-2.5$
shown in a color-coded plot against the normalized trap coordinates.
}
\label{fig:U24D16}
\end{figure}

Despite the increased asymmetry of the trap, the atomic distribution 
changes very little, and the pair correlation
exhibits similar oscillations as those shown in figure \ref{fig:U24D16}.
This is yet another evidence that the LDA is a rather 
good approximation in a system of 1D population-imbalanced fermions.
Recently the phase separation between the FFLO condensate and fully paired or
fully polarized wings of the trap has been studied in \cite{2010arXiv1001.4720H},
where a completely asymmetric setup, which corresponds to $r\rightarrow\infty$,
was used and compared with analytic results based on the combination
of the Bethe ansatz solution and LDA.

\begin{figure}
%\bc
\hspace{\fill}
\includegraphics[width=12cm]{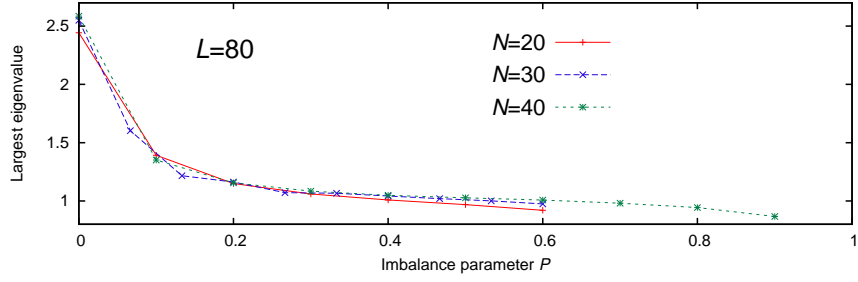}%{1DFFLO/cv1/Fig3/MaxEigVsP.eps}
%\ec
\caption{%(Color online)
Largest eigenvalue $\lambda^\mathrm{max}_2$ of the reduced two-body density matrix
plotted against the imbalance parameter $P$ for $N=20, 30$ and $40$.
The parameters are $L=80$, $A/J=1=6400/L^2$ and $U/J=-10=-800/L$. 
Reproduced from \cite{tezuka:110403} with modified color codes.
}
\label{fig:MaxEigVsP}
\end{figure}

\subsubsection{Two-body pair correlation matrix\\}
To quantify how much atoms contribute to the (quasi-) condensation in the
population-imbalanced 1D system,
we examine the behaviour of the eigenvalues of the two-body reduced density matrix,
from which we can identify the presence or absence of the off-diagonal
long-range order \cite{RevModPhys.34.694}.
We consider the two-body reduced density matrix
\be
\rho^\mathrm{full}_{\alpha \beta; \gamma \delta} \equiv
\langle \Psi_0|\hat c^\dag_\beta \hat c^\dag_\alpha \hat c_\gamma \hat c_\delta|\Psi_0 \rangle,
\ee
where $\alpha, \beta, \gamma, \delta$ are one of the $M$ possible states
that $N$ fermions in the system can each take.
As proved in \cite{RevModPhys.34.694}, the upper bound of the eigenvalue $\lambda_2$ of
$\rho^\mathrm{full}_{\alpha \beta; \gamma \delta}$
is $N(1-(N-2)/M)$, and 
the system possesses an off-diagonal long-range order
if and only if $\lambda_2$ is of the order of $N$.

Since pairing occurs between up-spin and down-spin atoms,
it suffices to calculate the
part of $\rho^\mathrm{full}$
for which $\alpha$ and $\beta$ as well as $\gamma$ and $\delta$
have different spins.
Therefore, we calculate
\be
\rho_{i_\uparrow, i_\downarrow; j_\uparrow, j_\downarrow}
\equiv \langle \Psi_0|\hat c^\dag_{i_\uparrow, \uparrow} \hat c^\dag_{i_\downarrow, \downarrow}
\hat c_{j_\uparrow, \uparrow} \hat c_{j_\downarrow, \downarrow}|\Psi_0\rangle.
\label{eqn:2BDM}
\ee
The eigenstates of this matrix with large eigenvalues correspond to states
with high pair concentration.
It can be shown that 
they become independent of the lattice discretization in the limit
of $L\rightarrow\infty$.
Note that while the dimension of the matrix is $L^2$, the trace of the matrix,
which is also the sum of the eigenvalues, equals $N_\uparrow N_\downarrow \ll L^2$,
and the maximum possible eigenvalue, which is close to $N_\downarrow$,
is realized when every spin-down atom is paired with a spin-up atom and when these pairs are
all condensed into a single state.

We have diagonalized the two-body density matrix for the ground state
that is obtained in the DMRG simulation for each parameter set.
Because the number of independent matrix elements scale as $\mathcal{O}(L^4)$ with increasing
the number of sites $L$, we used $L=80$ and limited $N$ up to $40$.
The number of states per DMRG block is also limited to $m=100$,
because for $L=80$, the energy and largest eigenvalues of the
two-body density matrix for the ground state converges for
smaller $m$ compared to the $L=200$ case.
The $i=i'$, $j=j'$ components of this matrix are the on-site pair correlation functions
and correspond to the pair amplitude of the condensation of on-site ``molecules.''
In Ref.~\cite{feiguin:220508}, the matrix of the on-site pair correlation,
\be
\rho^\mathrm{on-site}_{i; j}
= O_\mathrm{on-site}(z_i, z_j)
\equiv \langle \Psi_0|\hat c^\dag_{i, \uparrow} \hat c^\dag_{i, \downarrow}
\hat c_{j, \uparrow} \hat c_{j, \downarrow}|\Psi_0\rangle,
\ee
was diagonalized and the resulting eigenfunctions showed periodic oscillations.
A similar feature is confirmed in our analysis of the full two-body pair correlation matrix.

In figure \ref{fig:MaxEigVsP} we plot the largest eigenvalue of $\rho$ against the imbalance
parameter $P$ for different numbers of atoms $N$.
The dependences on $P$ for different values of $N$ are strikingly similar.
This indicates that, in the trapped system,
the lowest energy state of atom pairs is
occupied only by a few pairs, before $N$ reaches $20$,
and they do not accommodate additional pairs of atoms,
probably due to the effect of the trap potential which localizes the low energy states.

\begin{figure}
%\bc
\hspace{\fill}
\includegraphics[width=12cm]{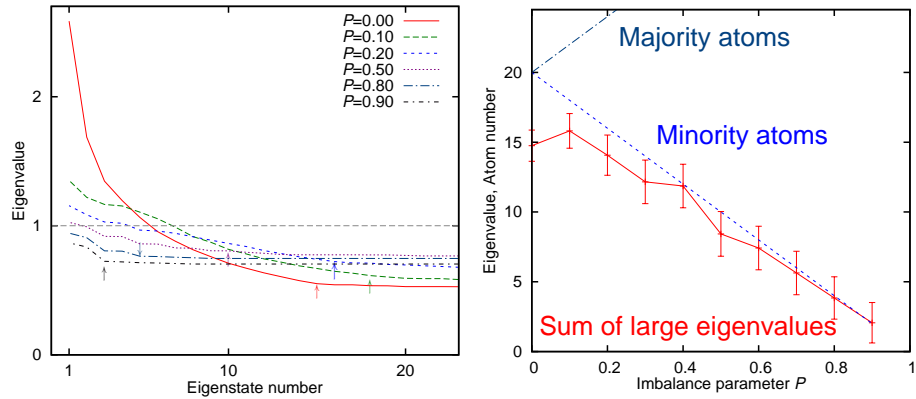}%{EigvVsOrd.eps}
%\ec
\caption{%(Color online)
Left column: Distribution of the $n$-th largest eigenvalues of the
two-body density matrix $\rho$ (\ref{eqn:2BDM})
plotted against $P$ for $N=40$.
The observed kinks in the distribution curves are marked with vertical arrows.
The parameters are $L=80$, $A/J=1=6400/L^2$ and $U/J=-10=-800/L$.
Right column: Numbers of majority and minority atoms, and the
sum of the eigenvalues of $\rho$ that are enhanced by the effect of
pair quasi-condenstaion,
plotted against $P$, for
$L=80$, $N=40$, $A/J=1=6400/L^2$ and $U/J=-10=-800/L$.
}
\label{fig:EigvDistr}
\end{figure}

In the left column of figure \ref{fig:EigvDistr} we plot the $n$-th largest eigenvalue of $\rho$
against $n$.
While the distribution is by definition a decreasing function of $n$,
we find a kink in the distribution at different values of $n$ for different $P$.
Without the interaction, the eigenvalues of the reduced density matrix are always
unity or zero, because each single-atom state is either filled or empty.
The kink in the distribution suggests that
the eigenvalues larger than the kink value are enhanced by the effect of pair
quasi-condensation.
In the right column of figure \ref{fig:EigvDistr} we plot the sum of such
enhanced eigenvalues against $P$.
The sum is close to the number of minority atoms in the system, which indicates
that almost all the minority atoms contribute to the quasi-condensation
and the oscillating pair correlation.

\section{Dynamics}
Now we study the dynamics of a 1D, harmonically trapped Fermi gas 
after a population imbalance is introduced
by flipping the spin of a single atom.
The flipping can be realized by applying a microwave radiation, whose energy
matches the hyperfine energy level separation.

\subsection{Formulation: Real-time DMRG simulation after a spin-flip excitation}
We simulate a sudden spin flip by applying a linear combination of local
spin raising operators $s^+_l$,
\be
\hat S^+ \equiv \sum_l c_l \hat s^+_l,
\ee
under spatial discretization.
We obtain the wavefunction of the system for the ground state in the $N_\uparrow=N_\downarrow=N$-spin sector
$|\Psi_0\rangle$ as discussed in the previous section,
and then apply $\hat S^+$ to obtain the normalized (with a constant $N$) wavefunction as the initial state of the
spin-flipped system,
\be
|\Psi(t=0)\rangle \equiv N S^+ |\Psi_0\rangle.
\label{eqn:spin-flip}
\ee

The spin flip does not in general map the ground state in the $(N_\uparrow, N_\downarrow) = (N, N)$ sector
to an eigenstate in the $(N_\uparrow, N_\downarrow) = (N+1, N-1)$ sector.
Therefore, the system will evolve in time, showing some real-time dynamics.
As a realistic case, we consider a situation in which the coefficient $c_l$ is spatially localized,
which corresponds to the local spin excitation.
This can be implemented experimentally by using a magnetic field gradient.

In the case of an attractively interacting system, the spin flip leads to
pair-breaking and
increases the energy of the system relative to the ground state in the $(N+1, N-1)$.
As discussed above, this ground state consists of a Larkin-Ovchinnikov condensate
at the trap center and an equal-population condensate near the trap fringes,
due to the low ($P=1/N<P_c$) spin polarization.

For the exact calculation (after the spatial discretization), the full Hilbert space has to be
considered, and the size of the Hilbert space rapidly increases.
For example, for $N=4$ in the $L=16$-site system,
the dimension of the Hilbert space is already $2446080$.
Several methods to apply DMRG to simulate real-time dynamics of quantum systems
have been implemented
\cite{PhysRevLett.88.256403, PhysRevLett.91.049701,
1742-5468-2004-04-P04005, PhysRevLett.93.076401,
PhysRevB.70.121302, manmana:269, PhysRevB.72.020404, PhysRevB.73.195304}.
Comparison of some of these methods have been conducted in
\cite{1367-2630-8-12-305}.
From the ground state in the restricted Hilbert space obtained in the DMRG simulation,
we calculate the spin-flipped state, and take this state as the state at $t=0$. 
Then we apply the time-stepping targeting method \cite{PhysRevB.72.020404} to simulate the real-time
evolution of this initial wavefunction $|\Psi(t=0)\rangle$.

In the time-stepping targeting method \cite{PhysRevB.72.020404}, the Schr\"odinger equation
\be
i\hbar\frac{\partial}{\partial t}|\Psi(t)\rangle
= \hat \mathcal{H}|\Psi(t)\rangle
\ee
is treated as the equation of motion for the many-body wavefunction $|\Psi(t)\rangle$,
and the Runge-Kutta method is used to obtain
$|\Psi(t+\Delta t)\rangle$, where $\Delta t$ is the time step,
within the Hilbert space selected in the previous finite-size system DMRG step.
At each step, $|\Psi(t)\rangle$, $|\Psi(t+\Delta t)\rangle$, and several other wavefunctions
$|\Psi(t+a_j\Delta t)\rangle$ are targeted in the choice of basis. After one iteration of the finite-system
algorithm DMRG, $t$ is increased by $\Delta t$ and the process is repeated.
Following \cite{PhysRevB.72.020404}, here we target four time steps
$|\Psi(t)\rangle$, $|\Psi(t+\Delta t/3)\rangle$, $|\Psi(t+2\Delta t/3)\rangle$, and $|\Psi(t+\Delta t)\rangle$,
whose weights in the density matrix are $1/3, 1/6, 1/6$ and $1/3$, respectively.
In this method, as in the methods based on the Suzuki-Trotter decomposition \cite{1742-5468-2004-04-P04005,PhysRevLett.93.076401},
the reduced Hilbert space in the subsystems after several DMRG iterations no longer contains
much of the information on $|\Psi(t=0)\rangle$.
Use of the Runge-Kutta method in simulating real-time dynamics has also been discussed
in \cite{1742-5468-2004-04-P04005}.

Here we note that the dynamics of the superfluid to Mott insulator transition
\cite{PhysRevA.70.043612, PhysRevLett.98.180601}, % Clark PRA 70, 043612 % Kollath PRL 98, 180601
spin-charge separation of a 1D Fermi gas \cite{PhysRevLett.95.176401, EPL.86.57006}, % Kollath % Ulbricht
dynamics of spinless fermions after an interaction quench \cite{PhysRevLett.98.210405, PhysRevB.79.155104}, % Manmana
expansion of a (bosonic or fermionic) Mott insulator \cite{1367-2630-8-8-169, PhysRevA.78.013620, PhysRevA.80.041603},
% K Rodriguez New J Phys 8 169 (2006) % Heidrich-Meisner PRA 78, 013620 % PRA 80, 041603R
relaxation of atoms after a superlattice is modified \cite{2008PhRvA..78c3608F,JPSJ.78.123002}, % Flesch PRA 78, 033608
quench dynamics of the Bose-Hubbard model \cite{1742-5468-2008-05-P05018}, % Lauchli J Stat Mech 2008 P05018
and damping of dipole oscillations of
a 1D Bose gas after a sudden displacement of the trap \cite{PhysRevLett.102.030407} % Danshita
have been studied by the time-dependent DMRG or
a related method, time-evolving block decimation (TEBD) \cite{PhysRevLett.91.147902}.
For a mixed state, the real-time evolution of the density matrix of the system can
be simulated in DMRG.
Simulations of the dynamics for finite-temperature spin systems have also been carried out
\cite{PhysRevB.71.241101, PhysRevB.79.245101}.
Here, we show the results of our simulation for the dynamics of pure states at $T=0$.

\begin{figure}
%\bc
\hspace{\fill}
\includegraphics[width=12cm]{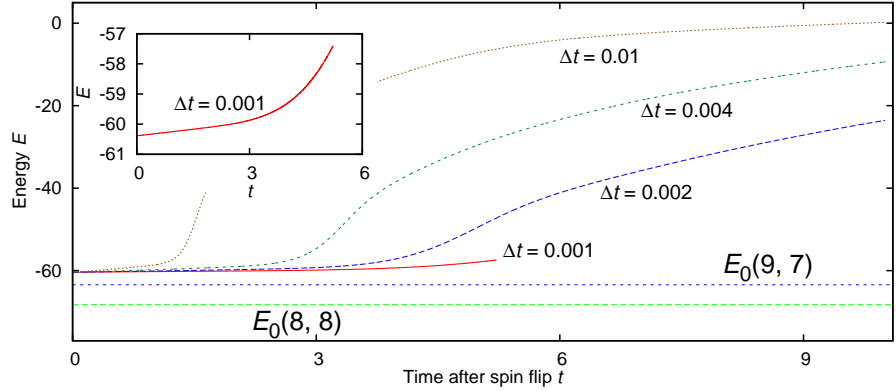}%{EnergyDltT.eps}
%\ec
\caption{%(Color online)
Energy of the simulated wave function after the spin-flip excitation at $t=0$
for $L=32$, $z_c=0$, $w=1/16$, $U/J=-8$ and $A/J=1$.
$E_0(N_\uparrow, N_\downarrow)$ denotes the ground-state energy in the sector
of $N_\uparrow$ spin-up atoms and $N_\downarrow$ spin-down atoms of the Hamiltonian
(\ref{eqn:Hamiltonian}). The curve for $\Delta t = 0.001$ is replotted in the inset.}
\label{fig:EnergyDltT}
\end{figure}

\subsection{Results: Diffusion of the excess majority atom and destroyed correlation}
In the following, a spin-flipping excitation is parameterized by two parameters,
the center of excitation $z_c$ and width $w$. The excitation factor for each site $l$, $s_l$,  is 
determined by
\be
s_l = C \exp{[-(z_l - z_c)^2/w^2]},
\ee
where $C$ is the normalization constant.

First, we examine the case in which the spin is flipped close to the trap center ($z_c=0$).
We consider the initial state obtained by flipping one of the down spins from the
ground state in the $(N_\uparrow, N_\downarrow) = (8,8)$ sector,
for $U=-8$ and $A=1$.
We take $L=32$ and $w=1/16$ and retain $m=300$ states per DMRG block.
Here,
$J = \hbar^2 L^2 / (8m_0 l^2) = 39.4~\mathrm{nK}\times k_\mathrm{B}$,
and we take $\hbar/J = 194~\mu\mathrm{s}$ as the unit of time.
In figure \ref{fig:EnergyDltT} we plot the evolution of the energy, defined by
\be
E(t) \equiv \langle \Psi(t)|\mathcal{H}|\Psi(t) \rangle,
\ee
where $|\Psi(t)\rangle$ is the calculated wave function at time $t$ after the spin flip at $t=0$.
We observe that the energy goes up rapidly for $\Delta t = 0.01$ after $t\simeq 1.3$.
Because the Hamiltonian is always the same, this increase in energy is
due to the accumulated error in the Runge-Kutta numerical integration.
For $\Delta t=0.002$, the increase of the energy occurs later and it is less significant.
For $\Delta t=0.001$, the energy is almost constant until $t\simeq 4$.
We present the result for $\Delta t = 0.001$ up to $t=4$.

\begin{figure}
%\bc
\hspace{\fill}
\includegraphics[width=12cm]{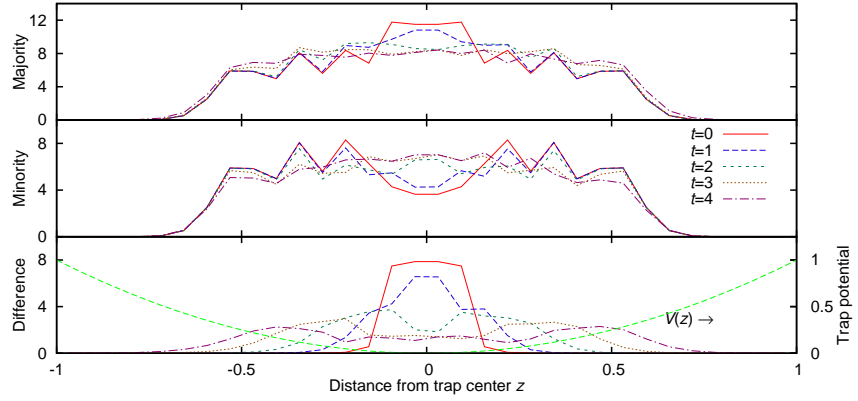}%{U-8tSite.eps}
%\ec
\caption{%(Color online)
Density distributions for different time steps after the spin-flip excitation at $t=0$
and the trap potential
plotted against the normalized coordinate, for $L=32$, $z_c=0$, $w=1/16$,
$U/J=-8$ and $A/J=V(\pm 1)/J=1$.}
\label{fig:Dyn0_NvsZ}
\end{figure}

For $U=-8$, we plot the density distribution for different time steps in figure \ref{fig:Dyn0_NvsZ}.
Because of the spin flip, initially (at $t=0$) the majority density is largest close to the trap center,
and the minority density has a deep dip there, making the density distribution localized there.
As the time elapses, the majority density decreases and the minority density increases
at the center, and the density difference exhibits two peaks at both sides of the central,
decreasing peak. Up to $t=4$, these two peaks move away from the trap center.
There remain three peaks of the density difference.

\begin{figure}
\bc
\hspace{\fill}
\includegraphics[width=10cm]{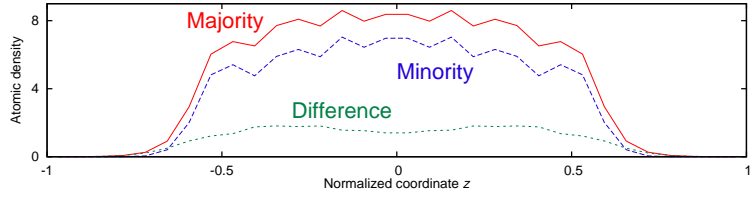}%{L32U9D7.eps}
\ec
\caption{%(Color online)
Density distributions for the ground state for $(N_\uparrow, N_\downarrow) = (9,7)$ atoms
in the trap with $L=32$, $A/J=1=1024/L^2$ and $U/J=-8 = -256/L$.}
\label{fig:density_9_7}
\end{figure}

This is to be compared with the density distribution for the $(N_\uparrow, N_\downarrow) = (9,7)$ sector
of the Hamiltonian (\ref{eqn:Hamiltonian}) (see figure \ref{fig:density_9_7}).
The majority and minority density profiles are not as smooth as those after the spin flip.
Such oscillations in the density profiles have also been observed for larger $L$;
they originate from the formation of tightly bound pairs,
rather than the lattice discretization
\cite{2008PhRvA..77e3614M}.
In the excited state after the spin flip,
such oscillations in the ground state
are smeared out because of the weaker binding between spins and the excess
kinetic energy.

In figure \ref{fig:density_9_7}, we see two peaks of the density difference
that correspond to the pair amplitude nodes of the LO condensate.
While the profile at $t=3$ after a spin-flip excitation
in figure \ref{fig:Dyn0_NvsZ} might seem similar,
the peak of the density difference continues to move outward
and the profile at $t=4$ has three peaks (including the one at $z\simeq -0.45$).
While the local spin excitation develops into an oscillating density difference
profile, the oscillation does not guarantee an LO condensate.

\begin{figure}
%\bc
\hspace{\fill}
\includegraphics[width=12cm]{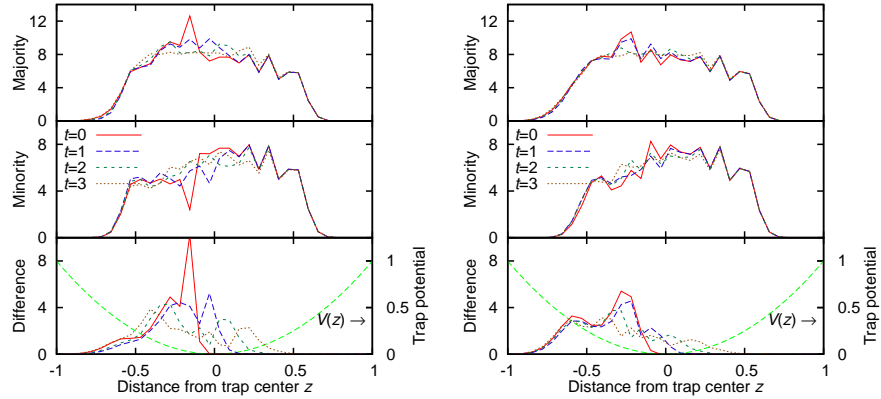}
%\ec
\caption{%(Color online)
Density distributions at different times after the spin-flip excitation
at $t=0$ with $z_c=-0.3125$ (left) and $z_c=-0.625$ (right),
plotted against the normalized coordinate for $L=32$, $w=1/16$,
$U/J=-8$ and $A/J=1$.}
\label{fig:Dyn5_10_NvsZ}
\end{figure}

Next we examine the effect of the location of the spin flip.
In figure \ref{fig:Dyn5_10_NvsZ},
we plot the density profile after the spin flip at two different $z_c$ off the center,
$z_c = -0.3125$ and $-0.625$, respectively.
The resulting density difference profile at $t=0$ does not have a peak at $z_c$,
because in calculating the excited state (\ref{eqn:spin-flip}), 
the amplitude of the spin-flipped wave function $\hat S_l^+ |\Psi_0\rangle$,
$\langle \Psi_0|\hat S_l^- \hat S_l^+ |\Psi_0\rangle$, depends on the site number $l$.
As the state evolves in time, the peak splits into two and they move to the left and right,
as in the case of the spin flip at $z=0$.
However, the movement of the density difference peak to the direction of $z>0$ is considerably
faster than that to the direction of $z<0$, as 
the kinetic energy of the atoms is greater for $z\sim 0$.

In figure \ref{fig:Dyn5_10_NvsZ}, as the spin flip location is removed from the trap center,
we observe that
the change of the minority population in the side of the spin flip ($z<0$)
becomes less pronounced, and that in the other side of the trap,
the densities of both components stay almost the same.
This is because the spin-imbalanced gas in the $z<0$ side pushes the
spin-balanced region ($z>0$) away, thus compensating for the reduced kinetic energy
by increasing the potential energy in the trap, but infiltrating this region only slowly.

Finally we look at the effect of the interaction strength.
For a smaller $|U|$, which corresponds to a smaller $|g^\mathrm{1D}|$ and 
a larger $|a_s^\mathrm{1D}|$, an increase in energy due to the spin flip is smaller,
but at the same time, the effect of Fermi pairing might become weaker.

\begin{figure}
%\bc
\hspace{\fill}
\includegraphics[width=12cm]{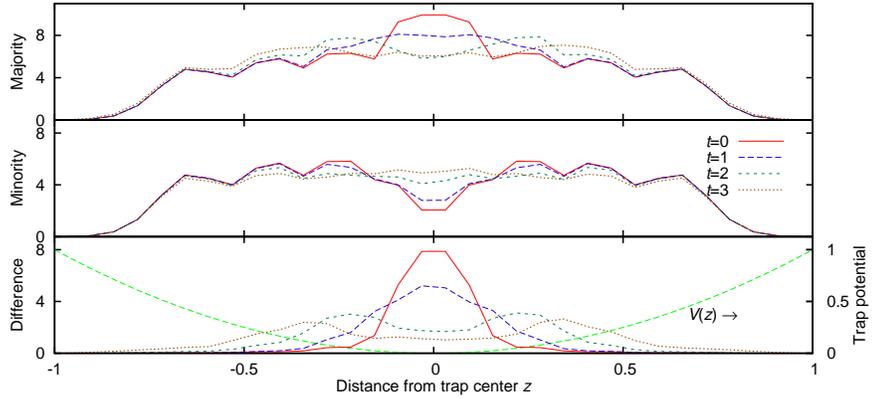}
%\ec
\caption{%(Color online)
Density distributions at different times after a spin-flip excitation at $t=0$
plotted against the normalized coordinate, for $L=32$, $z_c=0$, $w=1/16$,
$U/J=-2$ and $A/J=1$.}
\label{fig:U-2_Dyn0_NvsZ}
\end{figure}

In figure \ref{fig:U-2_Dyn0_NvsZ} we show the evolution of the system for $U/J=-2$ and
$A/J=1$, after the spin flip at the trap center ($z_c=0$).
Observe that, while the atoms are extended over a wider region due to the weaker
attraction, the speed of the movement of the two density-difference peaks is 
almost the same as in the case of $U/J=-8$.
The insensitivity of the speed to the interaction strength,
which is also observed for other spin flip parameters, suggests
that the momenta of the excess up-spin atoms are determined primarily
from the shape of the center peak whose width is independent of $|U|$,
and not much affected by $U$, which determines the pairing strength.
Indeed, if $|U|$ is kept constant and $w$ is increased, the
width of the initial distribution of the population imbalance increases,
and the speed of the diffusion of the density difference rapidly decreases;
this suggests that the speed may be determined from the uncertainty principle
in the initial peak where the excess up-spin atoms are introduced by the spin flip.

\section{Conclusion}

We have studied the ground state and dynamics of harmonically confined
1D Fermi gases with population imbalance.
We have discretized the 1D system and applied
the density-matrix renormalization group method,
which is numerically exact at least for the ground state in our setup,
and also allows the simulation of the dynamics of large quantum systems
at an unprecedented precision.
For the ground state,
we have demonstrated that for finite imbalance and attractive interaction,
a Larkin-Ovchinnikov type condensate forms at the trap center,
for a wide range of the polarization parameter up to almost unity.
Qualitatively this result is consistent with
both numerical studies \cite{feiguin:220508, 2008PhRvB..77x5105R, 2008PhRvA..77e3614M, 2008PhRvL.101l0404B, 2008PhRvA..78b3601L, batrouni:116405, 2008PhRvA..78c3607C},
and studies that utilize 
the exact solution for a uniform system together with
local density approximation
\cite{orso:070402, hu:070403, 2008PhRvA..77c3604G, 2009PhRvA..79d1603K, 2010JLTP..158...36Z}.

Beyond a critical value of $P$, the LO condensate
close to the trap center phase-separates from the spin-polarized Fermi gas
at the periphery,
regardless of the strength of interaction.
By studying the dependence of the eigenvalues of two-body density matrix
on $P$ and $N$, we have shown
that the oscillating pair correlation originates
from quasi-condensation of atom pairs, to which almost all the minority
atoms contribute.
The FFLO-like feature is found to be rather insensitive to the asymmetry,
which again supports the validity of the local density approximation.

We have studied
the dynamics after a spin-flip excitation by time-dependent
DMRG simulations.
Of particular interest is whether the density difference
oscillates as in the LO condensate in the ground state.
We have found that, while the snapshots of the density difference show
two or more peaks that originate from a single peak after the localized spin flip,
the peaks move in time and are not related with an FFLO-like state.

\section*{Acknowledgments}
Part of the computation has been done using the facilities of the Supercomputer Center, Institute for Solid State Physics, University of Tokyo. M.T. was supported by a Research Fellowship of the Japan Society for the Promotion of Science (JSPS) for Young Scientists.
\section*{References}
\bibliographystyle{unsrt_iop}
\bibliography{1DFFLO}

\end{document}